\documentclass[reprint,
superscriptaddress,
showpacs,preprintnumbers,
amsmath,amssymb,
aps,
prl,
]{revtex4-2}

\usepackage{bbm}
\usepackage{amssymb}
\usepackage{amsmath}
\usepackage{graphicx}
\usepackage{subfigure}
\usepackage[colorlinks,allcolors=blue]{hyperref}
\usepackage{epstopdf}
\usepackage{color}
\usepackage{bm}
\usepackage{mathrsfs}
\usepackage{dsfont}

\makeatletter

\newcommand{\Rmnum}[1]{\expandafter\@slowromancap\romannumeral #1@}

\makeatother
\begin{document}
\title{Witnessing Quantum Incompatibility Structures in High-Dimensional Multi-measurement Systems}
\author{Xiaolin Zhang}
    \affiliation{Ministry of Education Key Laboratory for Nonequilibrium Synthesis
and Modulation of Condensed Matter, Shaanxi Province Key Laboratory
of Quantum Information and Quantum Optoelectronic Devices, School of
Physics, Xi’an Jiaotong University, Xi’an 710049, China}
	\author{Rui Qu}
	\affiliation{Ministry of Education Key Laboratory for Nonequilibrium Synthesis
and Modulation of Condensed Matter, Shaanxi Province Key Laboratory
of Quantum Information and Quantum Optoelectronic Devices, School of
Physics, Xi’an Jiaotong University, Xi’an 710049, China}
\author{Zehong Chang}
	\affiliation{Ministry of Education Key Laboratory for Nonequilibrium Synthesis
and Modulation of Condensed Matter, Shaanxi Province Key Laboratory
of Quantum Information and Quantum Optoelectronic Devices, School of
Physics, Xi’an Jiaotong University, Xi’an 710049, China}
\author{Yunlong Wang}
\affiliation{Ministry of Education Key Laboratory for Nonequilibrium Synthesis
and Modulation of Condensed Matter, Shaanxi Province Key Laboratory
of Quantum Information and Quantum Optoelectronic Devices, School of
Physics, Xi’an Jiaotong University, Xi’an 710049, China}
\author{Zhenyu Guo}
\affiliation{Ministry of Education Key Laboratory for Nonequilibrium Synthesis
and Modulation of Condensed Matter, Shaanxi Province Key Laboratory
of Quantum Information and Quantum Optoelectronic Devices, School of
Physics, Xi’an Jiaotong University, Xi’an 710049, China}
	\author{Min An}
	\affiliation{Ministry of Education Key Laboratory for Nonequilibrium Synthesis
and Modulation of Condensed Matter, Shaanxi Province Key Laboratory
of Quantum Information and Quantum Optoelectronic Devices, School of
Physics, Xi’an Jiaotong University, Xi’an 710049, China}

	\author{Hong Gao}
	\affiliation{Ministry of Education Key Laboratory for Nonequilibrium Synthesis
and Modulation of Condensed Matter, Shaanxi Province Key Laboratory
of Quantum Information and Quantum Optoelectronic Devices, School of
Physics, Xi’an Jiaotong University, Xi’an 710049, China}
	\author{Fuli Li}
	\affiliation{Ministry of Education Key Laboratory for Nonequilibrium Synthesis
and Modulation of Condensed Matter, Shaanxi Province Key Laboratory
of Quantum Information and Quantum Optoelectronic Devices, School of
Physics, Xi’an Jiaotong University, Xi’an 710049, China}
	\author{Pei Zhang}
	\email{zhangpei@mail.ustc.edu.cn}
	\affiliation{Ministry of Education Key Laboratory for Nonequilibrium Synthesis
and Modulation of Condensed Matter, Shaanxi Province Key Laboratory
of Quantum Information and Quantum Optoelectronic Devices, School of
Physics, Xi’an Jiaotong University, Xi’an 710049, China}
\affiliation{State Key Laboratory of Applied Optics, Changchun Institute of Optics, Fine Mechanics and Physics, Chinese Academy of Sciences, Changchun 130033,China}
\date{\today}

\begin{abstract}
Quantum incompatibility, referred as the phenomenon that some quantum measurements cannot be performed simultaneously,
is necessary for various quantum information processing tasks, such as nonlocality and steering.
When these applications come to high-dimensional multi-measurement scenarios,
it is crucial and challenging to witness the incompatibility of measurements with complex structures.
To address this problem,
we propose a modified quantum state discrimination protocol that decomposes complex compatibility structures into pairwise ones
and employs noise robustness to bound incompatibility structures.
We then derive arithmetic bounds for arbitrary measurements and analytical bounds for mutually unbiased bases,
and capture some quantum incompatibility structures where measurements are partly compatible and partly incompatible.
Finally, we experimentally demonstrate our results
and connect them with quantum steering, quantum simulability and quantum communications.

\end{abstract}

\maketitle

\textit{Introduction.---}In quantum physics, some observables or measurements cannot commute with each other,
which implies that one cannot acquire precise information from them simultaneously \cite{robertson1929uncertainty}.
This feature is known as quantum incompatibility, which gives a notion of nonclassicality for quantum measurements or observables.
Quantum incompatibility is crucial to many applications,
including quantum cryptography \cite{gisin2002quantum},
contextuality \cite{amaral2018graph,liang2011specker,um2020randomness},
Einstein-Podolsky-Rosen (EPR) steering \cite{wiseman2007steering,uola2020quantum,qu2022robust,qu2022retrieving,PhysRevLett.132.210202},
Bell nonlocality \cite{brunner2014bell,hirsch2018quantum},
quantum computing \cite{mcnulty2023estimating,wang2021experimental},
and quantum communications \cite{guhne2023colloquium}.

When these applications come to high-dimensional multi-measurement scenarios (HDMMS),
the measurements, distributed probabilistically and defected due to imperfect experiment,
have complex incompatibility structures where measurements are partly compatible and partly incompatible
i.e., quantum incompatibility structures (QISs) \cite{guhne2023colloquium,heinosaari2016invitation,quintino2019device,PhysRevLett.131.120202}.
It is of fundamental and practical importance to witness QISs in HDMMS,
because they are necessary for producing a specific type of quantum correlation and appearing advantageous over classical communications.

\begin{figure}
    \centering
     \includegraphics[width=\linewidth]{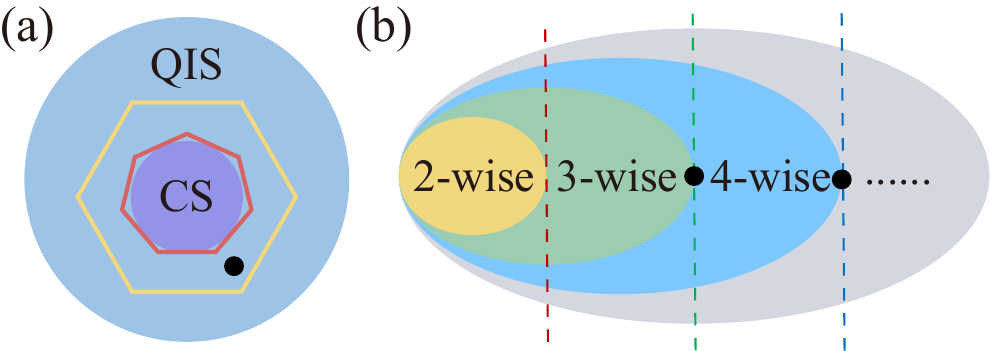}
     \caption{Geometrical interpretation of the contrast of Bell inequalities with our method to witness QISs and genuine $n$-wise incompatibility.
     (a)The scenario where measurements (the black dot) are detected incompatible by our method (outside the red polytope) but do not violate the Bell inequalities (inside the yellow polytope).
     (b)Witnessing genuine $n$-wise QISs with hyperplanes (the dashed lines), where the corresponding measurements are the tangent dots (the black dots).
     }\label{fig0}
 \end{figure}

Previous methods for witnessing QISs have relied on Bell nonlocality \cite{fine1982hidden,wolf2009measurements,quintino2019device,chen2016natural,chen2018exploring,quintino2016incompatible}.
However, these methods may be loose, as a violation of Bell's inequality implies incompatible measurements,
yet incompatible measurements do not necessarily result in Bell nonlocality.
For example, performing incompatible measurements on separable states cannot violate Bell inequalities (See Fig.~\ref{fig0}(a)).
Furthermore, it is tough to use Bell inequalities to witness QISs in HDMMS due to their fixed dimensions and measurement settings.
Recently, the quantum state discrimination protocol (QSDP) has been investigated as a witness for measurement incompatibility,
which is limited to pairs of measurements and cannot extend to QISs with complex structures in HDMMS
\cite{wu2021entanglement,uola2019quantifying,carmeli2019quantum,skrzypczyk2019all,zhang2022geometrical,carmeli2012informationally,carmeli2016verifying,carmeli2018state,skrzypczyk2019robustness,barnett2009quantum}.
Therefore, it is crucial and challenging to witness QISs in HDMMS.

In this work, we show how to modify the QSDP to certify a set of measurements respecting a given type of QIS by ruling out corresponding compatibility structures (CSs).
Our approach is based on the intuition that complex CSs can be decomposed by pairwise ones and a set of measurements violating specific CSs has limited robustness to noise.
Intuitively, our method constructs a polytope based on the targeted measurements with different probability distributions to witness QISs in HDMMS.
By adjusting hyperplanes of this polytope, our method is able to capture some specific QISs (See Fig.~\ref{fig0}(a)).

This article is organized as follows.
We begin by discussing the mathematical formulation of general CSs and genuine $n$-wise QISs with an example.
Next, we modify the QSDP and employ noise robustness to witness QISs in a semidefinite programming (SDP) formulation.
This approach derives arithmetic genuine $n$-wise QIS bounds for arbitrary measurements and general QIS bounds by constraining the weight of CSs.
We then discuss the case of symmetrical mutually unbiased bases (MUBs) and give their analytical genuine $n$-wise QIS bounds.
To demonstrate our results, we conduct an experiment to witness the genuine $3$ and $4$-wise QISs and a general QIS by adjusting hyperplanes (See Fig.~\ref{fig0}(b)).
Finally, we connect our results with EPR steering, quantum simulability and quantum communications.

\textit{General compatibility structures and genuine $n$-wise incompatibility.---}In quantum mechanics,
a general measurement $M$ is described by a set of positive operator-valued measurements (POVMs) with elements $M_{a}\succeq 0$,
where $\sum _{a}M_{a}=\mathbb{I}$ ($a=1,\cdots,o$ labels the outcome of the measurement and $\mathbb{I}$ is the identity operator).
A set of $m$ quantum measurements $\{M _{x}\}_{x=1}^{m}$, is said to be fully compatible if and only if
there is a valid parent POVM $\{G_{\lambda }\}$ such that
\begin{align}
    M_{a \vert x}=\sum_{\lambda}p(a \vert x,\lambda)G_{\lambda} \quad \forall a,x, \label{parent}
\end{align}
where $p(a \vert x,\lambda) \geq 0$ and $\sum_{a}p(a \vert x,\lambda)=1 \quad \forall x,\lambda$.
Otherwise, these measurements are incompatible.
In other words, a set of compatible measurements can always be implemented simultaneously by employing their parent measurement $\{G_{\lambda}\}$
and post-processing the results according to the conditional probability distributions $p(a \vert x,\lambda)$.

\begin{figure}
    \centering
     \includegraphics[width=\linewidth]{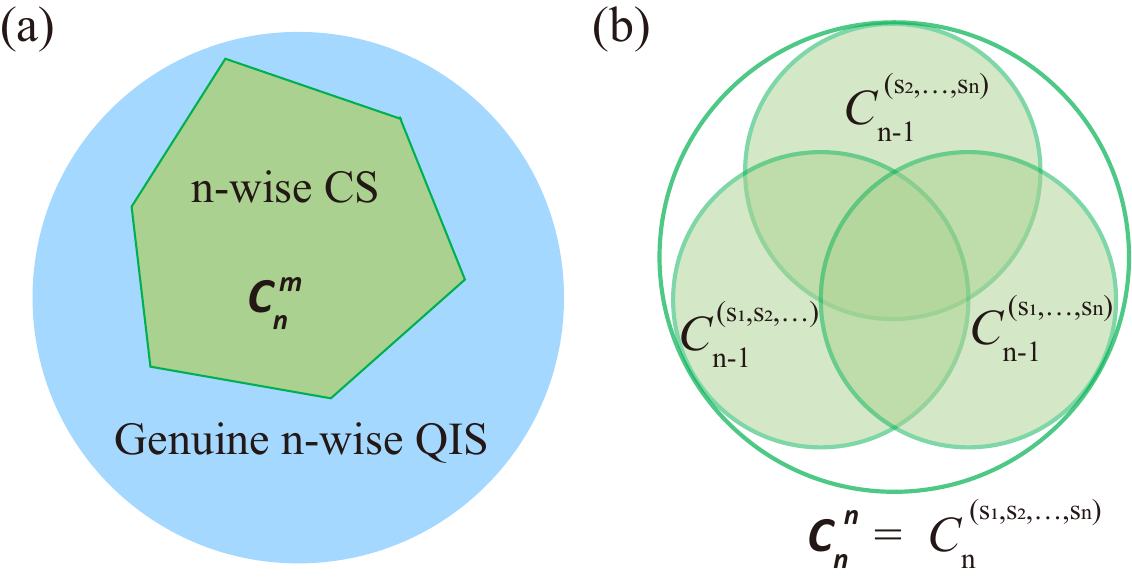}
     \caption{Geometrical interpretation of genuine $n$-wise QISs and $n$-wise CSs.
     (a)Genuine $n$-wise QISs characterized by the $\mathcal{C}_{n}^{m}$ polytope with $\frac{m!}{n!}$ facets $\mathcal{C}_{n}^{n}$.
     (b)One facet $\mathcal{C}_{n}^{n}$, which can be written as a convex combination of $(n-1)$-wise CSs.
     }\label{fig1}
 \end{figure}

When we consider the incompatibility of multiple measurements, partial notions of incompatibility may appear,
i.e., some measurements which are pairwise compatible, but not fully compatible as a complete set \cite{skrzypczyk2019all}.
In order to discuss this notion more specifically, we denote $\mathcal{C} = [C_{1},C_{2},\cdots,C_{N}]$ as general CSs and
$\{J_{a \vert x}^{C_{i}}\}$ as sets of measurements respecting the CS $\{C_{i}\}$,
where each $C_{i}$ denotes a CS of measurements.
For example, we denote $C_{1}=[{\{1,2\},\{1,3\},\{2,3\}}]$ and $C_{2}=[\{1,2,3\}]$, respectively.
Then, $C_{1}$ and $C_{2}$ indicates that measurements $1,2,3$ are pairwise compatible and $1,2,3$ are full compatible.
A set of measurements $\{M_{a \vert x}\}$ is $\mathcal{C}$-incompatible when it cannot be written as a convex combination of measurements that follow the CSs $[C_{1},C_{2},\cdots,C_{N}]$.
More precisely, a set of measurements $\{M_{a \vert x}\}$ is $\mathcal{C}$-incompatible if it cannot be written as
\begin{align}
    M_{a \vert x} = \sum_{i} p_{i}J_{a \vert x}^{C_{i}} \quad \forall a,x,\label{general compatibility}
\end{align}
where $\{p_{i}\}$ is the probability distribution of the CS $C_{i}$.
This approach allows for testing if there is a special type of CSs in a given set of measurements.

A set of measurements is genuinely $n$-wise incompatible when it cannot be written as a convex combination of measurements that are $(n-1)$-wise compatible on different partitions \cite{quintino2019device}.
It is called genuine incompatibility because it excludes all possible $2$-wise CSs that are the basic structures for quantum incompatibility.
Here we denote $\mathcal{C}_{n}^{m}$ as $n$-wise CSs in a set of $m$ measurements and $\mathcal{C}_{n}^{n}=[C_{n}^{(s_{1},s_{2},\dots,s_{n})}]$
as a specific $n$-wise CSs with measurements $S^{n}=(s_{1},s_{2},\dots,s_{n})$ in it.
Obviously, there are spontaneously $\frac{m!}{n!}$ groups of $n$ measurements in different $n$-wise CSs (See Fig.~\ref{fig1}(a)).
Therefore, a set of $n$ measurements is genuinely $n$-wise incompatible when it cannot be written as:
\begin{align}
    M_{a \vert x}&=J_{a \vert x}^{C_{n}^{(s_{1},s_{2},\dots,s_{n})}}= J_{a \vert x}^{C_{n}^{S^{n}}}=\sum_{i=1}^{n} p_{(S_{i}^{n})} J_{a \vert x}^{C_{n-1}^{S_{i}^{n}}}, \label{n compatibility}
\end{align}
where $S_{i}^{n}$ is an assemble from $S^{n}$ excluding $s_{i}$, i.e., $(s_{1},\dots,s_{i-1},s_{i+1},\dots,s_{n})$,
and $p_{(S_{i}^{n})}$ is the probability distribution of the corresponding group of measurements (See Fig.~\ref{fig1}(b)).

\begin{figure}
    \centering
    \includegraphics[width=\linewidth]{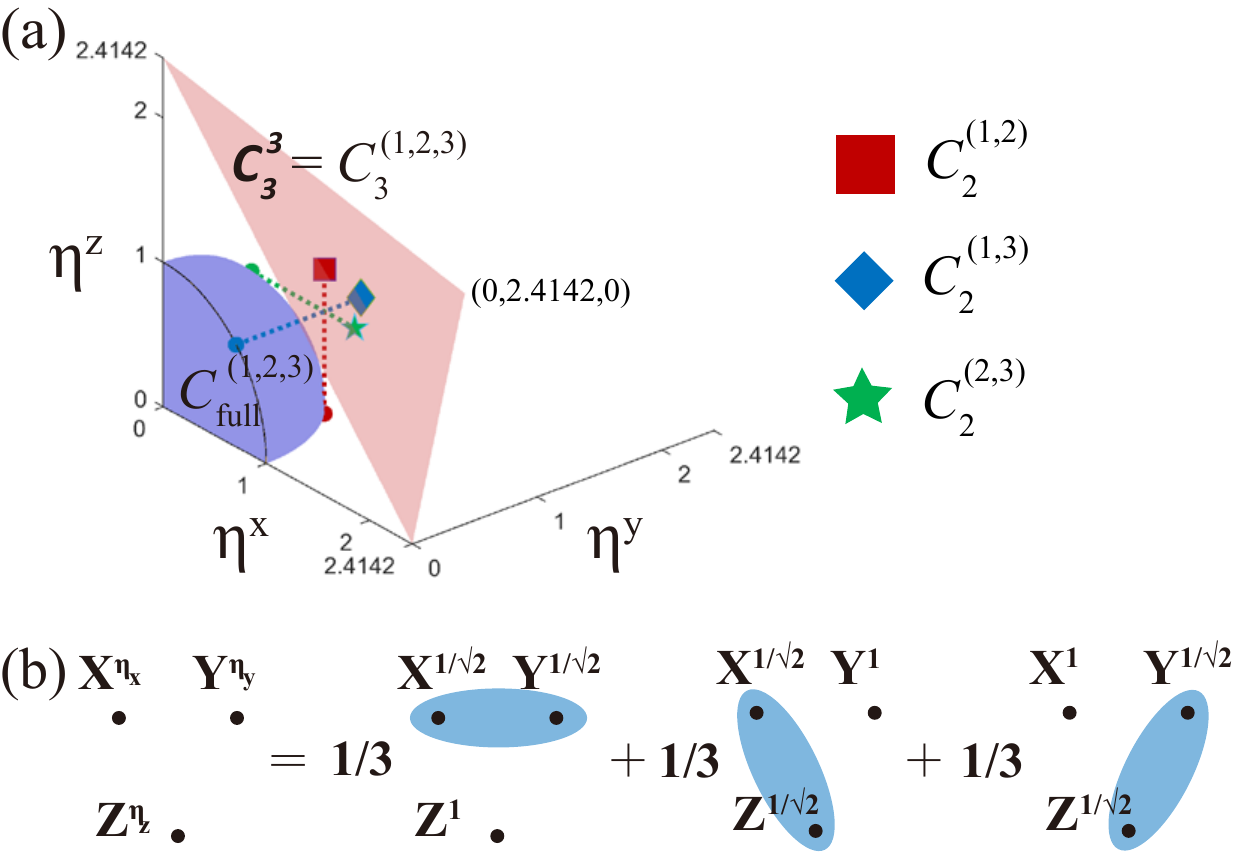}
    \caption{Geometrical interpretation of CSs of $3$ Pauli measurements in a qubit system.
    (a)CSs characterized by noise robustness,
    where three $2$-wise CSs $C_{2}^{(1,2)}$, $C_{2}^{(1,3)}$ and $C_{2}^{(2,3)}$ are tangent on the hyperplane $C_{3}^{(1,2,3)}$ with their projections on pair-wise compatibility.
    (b)The specific components of noisy Pauli measurements satisfying $\sum_{i}^{3} \eta_{i} = \sqrt{2}+1$.
    }\label{fig2}
\end{figure}

Here, we illustrate these concepts with an example of three noisy Pauli measurements in a qubit system:
\begin{align}
    M_{a \vert x}^{\eta}:=\eta_{x} \Pi_{a \vert x}+(1-\eta_{x})\frac{\mathbb{I}}{2},
\end{align}
where $x=1,2,3$ refers to each Pauli measurements $(X,Y,Z)$, respectively, $\Pi _{a \vert x}$ are their eigenprojectors, and $\eta_{x}$ is the corresponding sharpness.
Referring to Eq.~(\ref{parent}), these three measurements are fully compatible ($C_{full}^{(1,2,3)}$) when they have a common parent measurement for $\sum_{i}^{3} \eta_{i}^{2} \leq 1$ \cite{heinosaari2008notes},
and pairwise compatible ($[C_{2}^{(1,2)},C_{2}^{(1,3)},C_{2}^{(2,3)}]$) when two of them have a parent measurement for $\sum_{i}^{2} \eta_{i}^{2} \leq 1$ \cite{zhang2022geometrical,carmeli2019quantum}.
According to Eq.~(\ref{n compatibility}), this set of measurements can be written as a convex combination of sets where measurements are pairwise compatible for $\sum_{i}^{3} \eta_{i} \leq \sqrt{2}+1$ (See Fig.~\ref{fig2}):
\begin{align}
    M_{a \vert x} = p_{(1,2)}J_{a \vert x}^{C_{2}^{(1,2)}}+p_{(1,3)}J_{a \vert x}^{C_{2}^{(1,3)}}+p_{(2,3)}J_{a \vert x}^{C_{2}^{(2,3)}}. \label{Pauli convex}
\end{align}
Therefore, it is only for $\sum_{i}^{3} \eta_{i} > \sqrt{2}+1$ that they are genuinely $3$-wise incompatible.

Similar with the bound $\sqrt{2}+1$ for CS $\mathcal{C}_{3}^{3}$ of three noisy Pauli measurements,
the bound for $n$-wise CSs acts as a hyperplane for genuine $n$-wise QISs.
Any violation beyond this hyperplane will exclude all possible $(n-1)$-wise CSs.
Furthermore, $(n-1)$-wise CSs can always be decomposed by $(n-2)$-wise CSs,
and this decomposition continues for $(n-2)$-wise ones as well.

Using this approach, we can transform a complex CS into a convex combination of $2$-wise ones, which are the fundamental elements of compatibility
and rule out all possible CSs to witness genuine QISs.
Note that full CSs are the intersections of pairwise CSs and are automatically ruled out with pairwise ones.
To calculate the bound of genuine $n$-wise QISs,
we reformulate this problem into a modified QSDP within a SDP framework \cite{heinosaari2015noise,grant2014cvx,boyd2004convex}.

\textit{Witnessing genuine $n$-wise incompatibility in high-dimensional multi-measurement scenarios.---}To witness QISs in HDMMS,
we modify the QSDP to suit the concepts discussed above.

In the modified QSDP, Alice wants to witness the QISs of a set of measurements denoted as $\{ M_{a \vert x} \}_{a,x}$, (where $x = 1,\cdots ,m, a = 1,\cdots ,o)$,
and firstly she needs to share with Bob the probability distribution of measurements and operators $q(x)$ and $q(a \vert x)$, where $q(x)q(a \vert x)=q(a,x)$.

Then Bob prepares a set of ensembles $\{ \varepsilon _{x}\}_{x},x = 1,\cdots ,m$ with probability distribution $q(x)$ from Alice,
and sends quantum states $\varepsilon _{x}=\{ \rho _{a \vert x},q(a \vert x)\}_{x}$ with label $a$ and $x$ to her.
Upon receiving the state with its ensemble label $x$ and its corresponding probability distribution $q(x)$ from Bob,
Alice's task is to identify the label of the state, i.e.,
to correctly identify $a$.

To find the maximal guessing probability that the given measurements can achieve,
Alice performs $\{ M_{a \vert x} \}_{a,x}$ on the states from corresponding ensembles $\{ \varepsilon _{x}\}_{x}$, and evaluates the average probability of correctly guessing the state $a$ as:
\begin{align}
    P_{g}(\{ M_{ x} \}) =\max_{\{\varepsilon _{x}\}} \sum _{x,a} q(a,x)\operatorname{Tr}[\rho _{a \vert x}M_{a \vert x}], \label{first guessing}
\end{align}
where Bob optimizes his ensembles $\{\varepsilon _{x}\}$ after every round.

When Bob's ensembles are optimal,
Alice performs fixed measurements $\{G\}=\{G_{z}^{(s_{t},s_{r})}\}_{z}$ with its probability distributions $p(z)$ on every pair of ensembles $\{ \varepsilon _{s_{t}}\}$ and $\{\varepsilon _{s_{r}}\}$,
and evaluates the average correct guessing probability as:
\begin{align}\label{second guessing}
    P_{g}^{C}(\{ \varepsilon _{s_{t}}\},\{\varepsilon _{s_{r}}\})=&\max_{\{G\}} \sum_{x,a}^{x=s_{t},s_{r}} q(a,x) \times \\ \nonumber
    &\sum_{z}p(z) \operatorname{Tr}[\rho _{a \vert x}G_{z}^{(s_{t},s_{r})}]
\end{align}
where $\{G\}$ has the same effect with the parent measurement in Eq.~(\ref{parent}) and is optimized after every round to achieve the maximal probability that compatible measurements can achieve.

Since incompatible measurements can provide advantages over compatible measurements in QSDPs \cite{skrzypczyk2019all,carmeli2019witnessing},
a witness $W_{2}^{(s_{t},s_{r})}$ can be defined to detect the $2$-wise QIS ($C_{2}^{(s_{t},s_{r})}$) between measurements $M_{s_{t}}$ and $M_{s_{r}}$ as follows:
\begin{align}
    C_{2}^{(s_{t},s_{r})}:W_{2}^{(s_{t},s_{r})} =&  P_{g}^{x=s_{t},s_{r}}(\{ M_{ x} \})-P_{g}^{C}(\{ \varepsilon _{s_{t}}\},\{\varepsilon _{s_{r}}\})   \label{W2st}
\end{align}
where $P_{g}^{x=s_{t},s_{r}}(\{ M_{ x} \})$ represents the summation over $x=s_{t},s_{r}$,
and $M_{s_{t}}$ and $M_{ s_{r}}$ are compatible when $W_{2}^{(s_{t},s_{r})} \leq  0$, and incompatible otherwise.

In quantum technologies, measurement procedure with random measurement strategies, sequent measurement schemes or actual environmental disturbance,
can be considered as an unsharp or noisy version of corresponding measurements,
and thus we utilize noise robustness to form the hyperplanes to witness QISs.
By inserting noise into measurements with different probability distributions,
we propose a balanced noise robustness witness for $C_{2}^{(s_{t},s_{r})}$ in a SDP formulation~\cite{SM}\nocite{guo2023experimental,ecker2019overcoming,hu2020efficient}:
\begin{align}
	\text{maximize} \quad & R =  \sum_{x}q(x)\eta_{x} \nonumber\\
    s.t.\quad &  M_{a \vert x}^{\eta_{x}} = \eta_{x}M_{a \vert x}+(1-\eta_{x})\operatorname{Tr}[M_{a \vert x}]\frac{\mathbb{I}}{d} \nonumber \\
    & M_{a \vert x}^{\eta_{x}} = J_{a \vert x}^{C_{2}^{(s_{t},s_{r})}} \quad for \quad x=s_{t},s_{r},
\end{align}
where $\eta_{x}$ is the sharpness of the measurement $M_{ x}$,
the balanced noise robustness $R$ serves as a hyperplane separating CSs and QISs,
and $J_{a \vert x}^{C_{2}^{(s_{t},s_{r})}}$ is a set of measurements satisfies the compatibility case in Eq.~(\ref{W2st}).
After the optimization, these two measurements $M_{s_{t}}$ and $M_{s_{r}}$ are compatible when $R=1$, and incompatible when $R<1$.

Referring to the transition from $n$-wise CSs to $2$-wise ones,
we can obtain the mathematical form of hyperplanes for genuine $n$-wise QISs~\cite{SM}:
\begin{align}
	\max_{\{p\}} \quad &R = \sum_{x}q(x)\eta_{x} \nonumber\\
    s.t.\quad & M_{a \vert x}^{\eta_{x}} = J_{a \vert x}^{C_{n}^{(s_{1},s_{2},\dots,s_{n})}} \nonumber\\
    &= \sum_{i=1}^{n} p_{(S^{n})} (J_{a \vert x}^{C_{n-1}^{S^{n}_{i}}} + J_{a \vert x}^{\text{con}(s_{i})}) \nonumber\\
    &= \sum p_{(s_{t},s_{r})} (J_{a \vert x}^{C_{2}^{(s_{t},s_{r})}}+J_{a \vert x}^{con(S^{n}_{t,r})}), \label{hyperplane W}
\end{align}
where $\{p_{(s_{t},s_{r})}\}$ is the partial distribution for every $C_{2}^{(s_{t},s_{r})}$ and $\text{con}(x)$ represents constraints for the measurement $M_{x}$ to ensure the validity of the physics system.
Using this approach, a set of measurements are genuinely $n$-wise incompatible when $R<1$ and have CSs when $R=1$.

In this way, we establish a hyperplane based on balanced noise robustness $R$ for genuine $n$-wise QISs,
which excludes all possible CSs in a set of $n$ measurements.
We can also restrict some probability of CS to $0$ (for example $p_{(1,2,3)}=0$) before the optimization,
and the corresponding CS ($C_{3}^{(1,2,3)}$) may exist after surpassing the hyperplane.

In conclusion, by transforming the QIS problem into a SDP formulation,
we are able to find solutions numerically for arbitrary measurements with different probability distributions in HDMMS.
Furthermore, the proposed hyperplanes can form a well-defined geometrical polytope for
which many mathematical tools have been developed to study.

\textit{Bounding the genuine quantum incompatibility structures for mutually unbiased bases.---}MUBs are widely used in quantum information tasks and here we discuss their QIS bounds in HDMMS.

In a symmetrical case of a set of $m$ $d$-dimensional MUBs ($q(x)=1/n ,\quad q(a \vert x)=1/d \quad \forall a,x$),
there are $\frac{m!}{n!}$ groups of $n$ measurements.
Then the genuine $n$-wise QIS hyperplane for every group can be given~\cite{SM}:
\begin{align}
    \frac{1}{n}(\sum_{x}\eta_{x}) \leq \frac{\sqrt{d}-1}{n(d-1)}+\frac{n-1}{n},
\end{align}
which is analytical solution for genuine $n$-wise QISs for symmetrical MUBs in HDMMS.

Here we compare our method with Bell inequality.
When $d=2, n=3$, i.e. three Pauli measurements, the genuine $3$-wise QIS hyperplane is $\frac{1}{3}(\sum_{x}\eta_{x}) \leq \frac{\sqrt{2}+1}{3}$ by our method.
However, measurements respecting our genuine $3$-wise QIS hyperplane achieve $4.1213$,
which does not surpass the genuine QIS bound $4.9971$ derived by Bell inequalities~\cite{SM}.
This example demonstrates the advantages of our method in detecting QISs.

To demonstrate the feasibility to witness QISs in HDMMS,
we firstly discuss a symmetrical case where $d=3,m=4$ and $q(a \vert x)=1/3\quad \forall a,x$ for MUBs,
and there are four hyperplanes for the genuine $3$-wise QIS and one hyperplane for the genuine $4$-wise QIS.
These hyperplanes are respectively bounded by $\frac{\sqrt{3}+3}{6}$ and $\frac{\sqrt{3}+5}{8}$.

Then we discuss an asymmetric case where the probability distributions are set as $q(1)=\frac{1}{6},q(2)=\frac{1}{3},q(3)=\frac{1}{2}$.
The genuine $3$-wise QIS hyperplane is $\sum_{x}\eta_{x} \leq 0.864$ by Eq.~(\ref{hyperplane W}).
If we predetermine $p_{(1,2)}=0$ (which allows for $C_{2}^{(1,2)}$),
then the bound becomes $ 0.854$, which implies that we can capture the CS $C_{2}^{(1,2)}$ between the gap of $0.864$ and $0.854$.

In conclusion, we have given analytical genuine QIS bounds for symmetrical MUBs and compared them with those derived by Bell inequalities.
Then we have shown the feasibility to witness some general QIS where measurements have some CS but are incompatible as a set.
To show the feasibility and superiority of our method,
we next perform an experiment to demonstrate our results.

\begin{figure}
    \centering
    \includegraphics[width = 0.45 \textwidth]{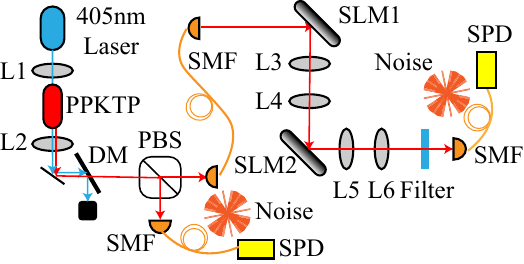}
    \caption{Experimental setup. Pumper, diode laser centered at 405~nm;
    L1-L6, lenses;
    PPKTP, periodically poled potassium titanyl phosphate;
    DM, dichroic mirror;
    PBS, polarizing beam splitter;
    SMF, single-mode fiber;
    HWP, half-wave plate;
    SLM, spatial light modulator;
    Filter, high-pass optical filter;
    Noise, variable-intensity LED;
    SPD, single-photon detector.
    }\label{Experiment}
\end{figure}

\textit{Experiments and results.---}We use photons carrying orbital angular momentum (OAM) degree of freedom to test the practicability of the modified QSD protocol, as depicted in Fig.~\ref{Experiment}.

We firstly utilize the type-\Rmnum{2} spontaneous parametric downconversion (SPDC) process to set up a heralded single-photon source,
which generates orthogonally polarized single photon pairs.
After a dichroic mirror (DM) removing the pump light, a polarization beam splitter (PBS) is set to separate the photons,
with horizontally polarized photons coupled into a single-mode fiber (SMF) for quantum prepare-and-measure scenarios,
while vertically polarized photons serve as a coincidence trigger.
After passing through the SMF,
the horizontally polarized photons are manipulated by the first spatial light modulator (SLM1) displayed with computer-generated holograms (CGHs) to prepare required quantum states \cite{bolduc2013exact}.
We use two 200~mm lenses L3 and L4 to constitute a $4f$ system, which helps image the plane of SLM1 to that of the second spatial light modulator (SLM2).
SLM2 is displayed with CGHs to measure the quantum states and the output photons are coupled to the SMF by a two 300~mm lenses L5 and L6.
Finally, we connect the two SMFs to two single-photon avalanche detectors (SPDs),
whose outputs are subsequently fed into a coincidence circuit with a coincidence time window of $1$ ns.

\begin{table}
    \caption{\label{tab:table1}
    Experimental results of genuine $3$ and $4$-wise QIS hyperplanes.
    }
    \begin{ruledtabular}
    \begin{tabular}{cccc}
       \multicolumn{4}{c}{Genuine symmetrical $3$ and $4$-wise QIS hyperplanes} \\ \hline
    \textrm{Hyperplanes}&
    \textrm{Predictions}&
    \textrm{Results}&
    \textrm{Planes}\\
    \colrule
    $\frac{1}{3}\eta_{1}+\frac{1}{3}\eta_{2}+\frac{1}{3}\eta_{3}$ &$0.789$  & $ 0.787 \pm 0.002$& $S_{1}$ \\
    $\frac{1}{3}\eta_{1}+\frac{1}{3}\eta_{2}+\frac{1}{3}\eta_{4}$ &$0.789$  & $0.787 \pm 0.002 $& $S_{2}$ \\
    $\frac{1}{3}\eta_{1}+\frac{1}{3}\eta_{3}+\frac{1}{3}\eta_{4}$ &$0.789$  & $ 0.787 \pm 0.002$ &$S_{3}$ \\
    $\frac{1}{3}\eta_{2}+\frac{1}{3}\eta_{3}+\frac{1}{3}\eta_{4}$ & $0.789$  &$0.787 \pm 0.002$ & $S_{4}$ \\
    $\frac{1}{4}\eta_{1}+\frac{1}{4}\eta_{2}+\frac{1}{4}\eta_{3}+\frac{1}{4}\eta_{4}$ &$0.842$& $0.841 \pm 0.001$ & $S_{5}$ \\ \hline
    \multicolumn{4}{c}{Genuine asymmetrical $3$-wise QIS hyperplane} \\ \hline
    $\frac{1}{6}\eta_{1}+\frac{1}{3}\eta_{2}+\frac{1}{2}\eta_{3}$&$0.864$ & $0.863 \pm 0.002$ & $S_{6}$ \\ \hline
        \multicolumn{4}{c}{General asymmetrical $3$-wise QIS hyperplane} \\ \hline
        $\frac{1}{6}\eta_{1}+\frac{1}{3}\eta_{2}+\frac{1}{2}\eta_{3}$&$0.854$ & $0.852 \pm 0.003$ &$S_{7}$\\
    \end{tabular}
    \end{ruledtabular}
    \end{table}

Specifically, we choose $l=-2,0,2$ modes to construct a set of four MUBs in a qutrit system,
and it is worth noting that coherent light cannot be used to replace the single-photon source,
because it cannot exactly generate the required single party qutrit states in the QSDP.

Next we explain how to perform noisy MUB measurements using actual noise.
We put two independent light sources after SLMs to ensure the noise is not affected by the mode generated.
The light sources are composed of two variable-intensity LEDs and are set before two SPDs, respectively,
to realize the independent adjustment of the noise proportions entering each single photon detector.
This is done by changing the brightness of the LEDs to control the total number of photons scattered into the detectors.

\begin{figure}
    \centering
    \includegraphics[width = 0.9\linewidth]{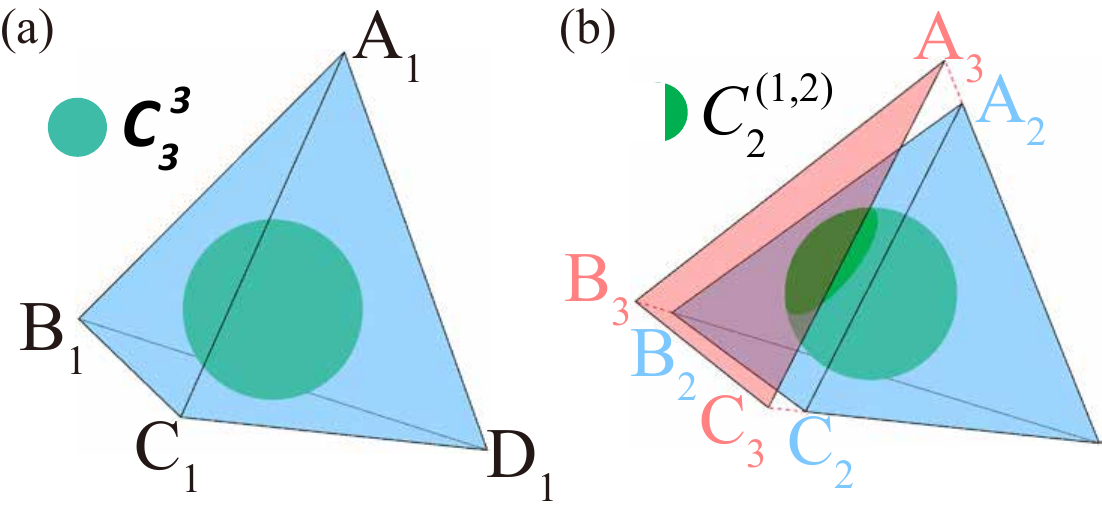}
    \caption{Simple geometrical illustration for results in Table~\ref{tab:table1}.
    (a)Symmetrical genuine $3$-wise QIS hyperplane, where the planes $(A_{1}B_{1}C_{1}),(A_{1}B_{1}D_{1}),(A_{1}C_{1}D_{1}),(B_{1}C_{1}D_{1})$ are $S_{1},S_{2},S_{3},S_{4}$, respectively.
    (b)Asymmetrical genuine $3$-wise QIS hyperplane, where the green part between $(A_{2}B_{2}C_{2})$ and $(A_{3}B_{3}C_{3})$ reveals that we capture the CS $C_{2}^{(1,2)}$ between the gap of $S_{7}$ and $S_{6}$.
    }\label{G3}
\end{figure}

We test the fidelity of the eigenstates of four MUBs and the average of fidelity is $F = 0.9873 \pm 0.0004 $~\cite{SM}.
Table~\ref{tab:table1} displays our experimental results and Figure.~\ref{G3} displays a geometrical illustration for results in Table~\ref{tab:table1}.
Note that, the $n$-wise CS is an irregular geometry as shown in Fig.~\ref{fig1}(a),
and in Figure.~\ref{G3} we use a sphere as a simplified model to represent the convex hull of CSs.

Our experimental results closely match the theoretical predictions,
which demonstrates the feasibility of our method for witnessing QISs in HDMMS.
From the respective of resource theories,
QIS can be seen as a kind of resources which is necessary for various quantum information tasks
and we witness the QIS resources in HDMMS.

\textit{Conclusions.---}In this work,
we have discussed the problem of witnessing QISs in HDMMS.
We derive genuine $n$-wise QIS arithmetic bounds for arbitrary measurements and analytical bounds for symmetrical MUBs using a modified QSDP.
Additionally, we experimentally test genuine $3$-wise and $4$-wise QIS in a qutrit system without relying on entangled sources,
and witness a general QIS where one CS $C_{2}^{(1,2)}$ is captured with other CSs ruled out.
Overall, our experimental results demonstrate that our approach offers a direct and intuitive tool to witness QISs in HDMMS,
and characterize the QIS resources, i.e., the hierarchy of quantum incompatibility resources.

Some open questions follow from our work.
First, if measurements are genuinely $\mathcal{C}$-incompatible and the state is $d$-Schmidt-rank pure entangled,
the EPR assemblage is genuinely $\mathcal{C}$-steerable \cite{quintino2019device}.
We have derived bounds for genuine QISs, and one can discuss the relation between genuine QISs and genuine $\mathcal{C}$ steering and perform an experiment.
Second, quantum incompatibility is seen as a specific case of simulability of quantum measurements,
which is an important way to compress quantum information and related to EPR steering \cite{PhysRevLett.129.190401}.
Referring to our work, one can consider quantum simulability structures where measurements are partially simulable and partially unsimulable.
Finally, in quantum communications, a pair of incompatible measurements show advantageous over compatible measurements in quantum random access code,
and one can consider the advantages of QISs over CSs in multi-measurement quantum random access code \cite{carmeli2020quantum}.

\begin{acknowledgments}
    This work was supported by the National Natural Science Foundation of China (Grant No. 12174301), the State Key Laboratory of Applied Optics and
    the Natural Science Basic Research Program of Shaanxi (Program No. 2023-JC-JQ-01).
\end{acknowledgments}

\bibliography{reference}

\end{document}